\documentclass[preprint]{elsarticle}
\usepackage{bm}
\usepackage[cp1251]{inputenc}
\usepackage[english]{babel}
\usepackage{amsfonts,amssymb,amsmath}
\usepackage{amsmath,graphicx}
\usepackage{graphicx,pifont} 
\usepackage{epsfig}

\begin{document}
\title{Relativistic kinetic theory: toward the microscopic substantiation of zeroth law of thermodynamics}

\author{A. Yu. Zakharov~$ ^{1} $,  V. V. Zubkov~$ ^{2} $}

\address{$ ^{1} $ Yaroslav-the-Wise Novgorod State University, Veliky Novgorod, 173003, Russia}

\address{
	$ ^{2} $ Tver State University, Tver, 170002, Russia}

\ead{Anatoly.Zakharov@novsu.ru; victor.v.zubkov@gmail.com}

\begin{abstract}
	An exact closed relativistic kinetic equation is derived for a system of identical classical particles interacting with each other through a scalar field. The microscopic deterministic mechanism of the irreversible equilibration process in a relativistic classical system of interacting particles has been established. 
\end{abstract}

	\maketitle

\section{Introduction}
In the late 19th and early 20th centuries, there were two main concepts regarding the microscopic origin of the laws of thermodynamics. The first of these was developed in the classic works of Maxwell~\cite{Maxwell-1,Maxwell-23,Maxwell-4}, Boltzmann~\cite{Boltzmann1, Boltzmann2}, Gibbs~\cite{Gibbs} and Poincar\'e~\cite{Poincare} and was based on molecular kinetic concepts.
At that time, classical Newtonian mechanics was the only theory that allowed the description of the microscopic dynamics of many-particle systems. However, the reversibility of the equations of classical dynamics and the irreversible behavior of real systems are in a certain contradiction with each other. This contradiction was one of the reasons for the popularity of the works of Ostwald~\cite{Ostwald}, Mach~\cite{Mach}, Duhem~\cite{Duhem} and Helm~\cite {Helm}, based on the concept of energetism. Within the framework of this concept, the fundamental principle of all phenomena in the world is energy --- a kind of indestructible hidden substance. In particular, matter is just one of the manifestations of energy. Within the framework of the energetism  concept, the molecular-kinetic theory was interpreted as a primitive mechanistic picture, and the existence of atoms was actively questioned.

The fierce fight between the representatives of these concepts~\cite {Brush} almost stopped after the experimental proof of the existence of atoms and the main method of theoretical study of the equilibrium properties of matter became Gibbs' statistical mechanics, which is a integration of Newtonian classical mechanics in the Hamiltonian form and the concepts of the theory of probability in the form of probabilistic measures in the phase spaces of the systems under study. According to statistical mechanics, the relationship between the Hamiltonian of a system and its thermodynamic properties is realized through these probabilistic measures. The implementation of this connection for the purpose of calculating and analyzing thermodynamic functions is a problem of huge mathematical complexity. But the main problems of statistical mechanics are of a much more fundamental nature and are not at all reduced to computational problems.

\begin{enumerate}
	\item  	First of all, this is the problem of explaining and substantiating the zeroth principle of thermodynamics, according to which any isolated many-particle system, regardless of the initial state, irreversibly passes over time to a state with steady-state macroscopic parameters. The zeroth law of thermodynamics is postulated in both phenomenological thermodynamics and statistical mechanics. Note that the irreversibility of the dynamics of many-particle systems is only a necessary, but not a sufficient condition for the fulfillment of the zeroth principle of thermodynamics. From this it follows that the possibility of microscopic substantiation of the zeroth principle of thermodynamics in the framework of classical Newtonian mechanics is at least doubtful, because it is required to deduce macroscopic irreversibility from microscopic reversibility. In essence, both the concept of energetism and the concept of probability are not entirely correct attempts to go beyond the limits of classical mechanics. The doubtfulness of the first of them is due to the introduction of some fantastic hidden substance, and the second is due to the internal inconsistency between deterministic classical mechanics and probabilistic assumptions.
	\item The problem of internal inconsistency of the statistical approach to problems of classical mechanics. The fact of the matter is not only in the long-known paradoxes (in fact, internal contradictions) of Loschmidt and Zermelo that have not yet been rationally resolved. As shown in the works of Kac~\cite{Kac1, Kac2}, the use of probabilistic assumptions such as the molecular chaos hypothesis in the framework of the Kac dynamic ring model leads to a result that contradicts the exact dynamic solution of this model. This counterexample to statistical mechanics raises strong doubts about the internal consistency of statistical mechanics~\footnote{Note that the chaotic nature of trajectories in classical mechanics, caused by the instability of solutions of dynamic equations with respect to variations in initial conditions (for example, Sinai's billiards), does not solve the problem of microscopic substantiation of the laws of phenomenological thermodynamics.}.
	
\end{enumerate}

In the post-Gibbsian time, the development of classical (i.e., non-quantum) statistical mechanics was mainly focused on the following directions.

\begin{enumerate}
		\item Investigation of the mathematical structure of Gibbs' statistical mechanics (reduction of the problem of the partition functions calculating to systems of BBGKY-type equations for distribution functions~\cite{Bogoliubov,Bogoliubov_2,Born,Kirkwood,Yvon}, conditions for the existence of the thermodynamic limit~\cite{Dob-1,Fish-1,Fisher-Rue,Ruelle}, exactly solvable models~\cite{Baxter, Sutherland, Thompson, Minlos}, ergodic theory~\cite{Cornfeld}, etc.).
	
	\item Non-equilibrium statistical mechanics (kinetic equations of BBGKY type~\cite{Bogoliubov, Born, Kirkwood, Yvon, Prigogine}, method of non-equilibrium statistical operator~\cite{Zubarev1, Zubarev2, Zubarev3}, methods of the theory of random processes~\cite{Liggett, Attard} ).
	
	\item Relativistic generalization of the classical kinetic theory of gases~\cite{Juttner1, Juttner2, Tetrode, Fokker, Synge, Chernikov, Kuzmenkov1, Groot2, Cerc1, Liboff}, statistical mechanics~\cite{Balescu, Schieve, Hakim, Tolman, Haar, Nakamura, Quevedo, Lusanna}.
\end{enumerate}

Despite the outstanding achievements in theoretical studies of condensed matter in the framework of statistical mechanics, the fundamental problem of a consistent microscopic substantiation of the laws of thermodynamics remains unsolved. Because of the fundamental contradiction between classical Newtonian mechanics and the undoubted laws of thermodynamics, the microscopic substantiation of thermodynamics should be sought outside of classical mechanics.

In this regard, we note the works~\cite{Synge1, Driver1, Hsing, Driver2}, in which the dynamics of two-particle systems with retarded interactions between particles is investigated and it is established that the retardation of interactions leads to a radical rearrangement of the dynamics of the systems. A possible connection between the rearrangement of dynamics and thermodynamic effects was not indicated in these works.

In the works~\cite{Zakharov16-2, Zakharov19-1} it is shown that the retardation of interactions leads to the irreversibility of the dynamics of both many-particle and low-particle systems, and it is also hypothesized that the retardation of interactions between particles is a microscopic probability-free mechanism leading to thermodynamic behavior of the systems.

It is essential that the retarded interaction of particles cannot be described in terms of the potential energy, which depends on the simultaneous values of the coordinates of these particles. Therefore, the Hamiltonian of such a system as a function of instantaneous coordinates and momenta of particles also does not exist. This is in full agreement with the well-known result on the absence of a relativistically invariant Hamiltonian for a system of interacting particles~\cite{Landau, Currie1, Currie2} and at the same time indicates the need to search for an alternative concept of microscopic substantiation of thermodynamics.

Thus, the dynamics of a system of particles with retarded interactions should be described within the framework of field theory. The complete system of equations for the dynamics of the closed system `` particles + field '' consists of equations for the dynamics of particles and equations for the evolution of the field generated by these particles. In the absence of external fields, the field variables can be eliminated from this complete system of equations and a closed system of functional differential equations describing the particle dynamics can be obtained. As applied to a system consisting of point charged particles, this program was implemented in~\cite{Zakharov20, Zubkov21}.

Note that the evolution of a system of particles with instantaneous interactions between them can also be described in the framework of the field theory with an infinite velocity of propagation of interactions and with the subsequent elimination of field variables. However, there is a fundamental difference between systems with instant and retarded interactions, which manifests itself after the elimination of field variables. It is as follows.

\begin{itemize}

	\item The system with \textit{instantaneous} interactions is Hamiltonian one and, as a consequence, it contains Poincar\'e integral invariants (including the invariance of the Liouville phase volume), the Poincar\'e recurrence theorem holds, invariance with respect to the time reversal $ t \longrightarrow -t $, as well as existence and uniqueness theorems for the Cauchy problem.
	
	\item The system with \textit{retarded} interactions between particles, on the contrary, is not Hamiltonian and therefore Poincar\'e's invariants do not exist in it, Liouville's theorem and equation do not hold, the dynamics equations are not invariant with respect to time reversal. However, it is these properties of a system with retarded interactions that are consistent with the laws of thermodynamics and therefore can be used for a consistent microscopic substantiation of thermodynamics.

\end{itemize}

In recent works~\cite{Zakharov16-2, Zakharov17, Zakharov18, Zakharov19-1, Zakharov19-2}, the description of the dynamics of systems with retarded interactions was carried out in terms of microscopic (i.e. not-averaged) distribution functions
\begin{equation}\label{f(r,v,t)}
	f\left(\mathbf{r},\mathbf{v},t\right)  = \sum_{s}\, \delta\left( \mathbf{r}- \mathbf{R}_{s}\left( t\right) \right) \, \delta\left( \mathbf{v}- \mathbf{\dot{R} }_{s}\left( t\right) \right),
\end{equation}    
where $ \mathbf{R} _ {s} \left(t \right) $ is the radius vector of the $ s $-th particle depending on the time $ t $. It is shown that the retardation of interactions leads to the irreversibility of the dynamics of the system, and this property is equally valid both for many-body and for few-body systems.

In order to find sufficient conditions for equilibration, in~\cite{Zakharov20-1, Zakharov21-1} the dynamics of a one-dimensional chain of atoms with retarded interactions between them was investigated.

It is shown in~\cite{Zakharov20-1} that all free vibrations of a chain of atoms with retarded interactions are damped and at $ t \to \infty $ the system goes into a state of rest. All the kinetic energy of atoms is irreversibly completely transferred into the energy of the field, through which the interaction between the particles takes place.

In ~\cite{Zakharov21-1}, the dynamics of forced (under the action of an alternating external field) vibrations of a one-dimensional chain of atoms with retarded interactions is investigated. It is shown that stationary oscillations are established in the system as $ t \to \infty $. This transition is interpreted as the process of equilibration between a system of particles and an alternating external field.

The exact results of works~\cite{Zakharov20-1, Zakharov21-1} demonstrate a deterministic dynamic microscopic mechanism of thermodynamic equilibration for a simple particular model --- in a one-dimensional chain of atoms.

The purpose of this work is as follows.
\begin{enumerate}
	\item Creation of the classical relativistic dynamical theory of a system of identical particles interacting with each other through a scalar field of general form.
	\item  Determination of microscopic relativistic mechanisms of thermodynamical equilibration processes.
\end{enumerate}

\section{Relativistic kinetic equation}

As is known~\cite{Kosyakov,Poincare2}, the relativistic equation of motion of a particle has the form  
\begin{equation} \label{eq-motion} 
\frac{dp_{a}^{\mu } }{d\tau _{a} } =F_{a}^{\mu } \left(x_{a},\, p_{a} \right).  
\end{equation} 
Here $d\tau _{a} $  is proper time along the world line of a $a$-th particle, $x^{\mu } =\left(x^{0}, \mathbf{r}\right)=\left(ct,\mathbf{r} \right)$ is the space-time four-vector, $p_{a}^{\mu } =\gamma _{a} m_{a} \left(c,\mathbf{v}_{a} \right)$ is the energy-momentum four-vector for $a$-th particle, $F_{a}^{\mu } =\left(F_{a}^{0},\mathbf{F}_{a} \right)=\gamma _{a} \left( \frac{\left( \mathbf{v}_{a} \cdot  \mathbf{F}_{a} \right) }{c}, \mathbf{F}_{a} \right)$ is the four-vector of force, acting on $a$-th particle, $\gamma _{a} = \frac{1}{\sqrt{1-\left(\mathbf{v}_{a}/c\right)^{2}}} $ is the Lorentz factor.

After multiplying equation~\eqref{eq-motion} by the delta-functions in the four-dimensional Minkowski space-time $\delta ^{4}\! \left(x-x_{a} \left(\tau _{a} \right)\right)\delta ^{4}\! \left(p-p_{a} \left(\tau _{a} \right)\right)$, integrating over the proper time $\tau _{a} $ and summing over $ a $, we get
\begin{equation} \label{sum-a1} 
\begin{array}{c} 
	{\displaystyle \sum _{a}\int d\tau _{a} \ \frac{dp_{a}^{\mu } }{d\tau _{a} }\  \delta ^{4}\! \left( x-x_{a} \left(\tau _{a} \right) \right)\,  \delta ^{4}\!  \left( p-p_{a} \left(\tau _{a} \right) \right)  } \\ 
	{\displaystyle =\sum _{a}\int d\tau _{a}\,  F_{a}^{\mu }\,  \left(x_{a} ,p_{a} \right) \delta ^{4}\! \left( x-x_{a} \left(\tau _{a} \right) \right)\,  \delta ^{4}\!  \left( p-p_{a} \left(\tau _{a} \right) \right). } 
\end{array} 
\end{equation} 

Consider the microscopic distribution function for particles of type~$A$~\footnote{As indicated in~\cite{Klimontovich-1960}, the idea of introducing such an invariant construction~\eqref{FA(x,p)} belongs to R.L.~Stratonovich}:
\begin{equation} \label{FA(x,p)} 
{\mathcal F}_{A} \left( x,p \right)=\sum _{a}\int\, d\tau _{aA}\, \delta^{4}\! \left(  x-x_{aA} \left(\tau_{aA} \right)\right)  \, \delta^{4} \!  \left( p-p_{aA} \left(\tau _{aA} \right) \right).
\end{equation} 
Using the expression
\begin{equation} \label{sum-a2} 
\begin{array}{c} 
{\displaystyle \frac{d }{d\tau _{a} } \left[ \delta^{4}\! \left(  x-x_{aA} \left(\tau_{aA} \right)\right)  \, \delta^{4} \!  \left( p-p_{aA} \left(\tau _{aA} \right) \right) \right] } \\ \\ {\displaystyle =-\frac{p_a^{\mu } }{m_{a} } \left[ \frac{\partial }{\partial x^{\mu } }\delta^{4}\! \left(  x-x_{aA} \left(\tau_{aA} \right)\right) \right] \delta^{4} \!  \left( p-p_{aA} \left(\tau _{aA} \right) \right)}\\ \\
{\displaystyle -\frac{dp_{a}^{\mu } }{d\tau _{a} }\,  \delta^{4}\! \left(  x-x_{aA} \left(\tau_{aA} \right)\right) \left[  \frac{\partial}{\partial p^{\mu}}\delta^{4} \!  \left( p-p_{aA} \left(\tau _{aA} \right) \right)\right]  } 
\end{array} 
\end{equation} 
and integrating  by parts on left-hand side of \eqref{sum-a1} leads  to the relativistic kinetic equation in the {covariant} form:
\begin{equation} \label{dF-dx} 
\left( \frac{p^{\nu } }{m_{A} } \, \frac{\partial }{\partial x^{\nu } } + F^{\nu } \left(x,p\right)\, \frac{\partial  }{\partial p^{\nu } } + \frac{\partial  F^{\nu } \left(x,p\right) }{\partial p^{\nu } } \right)  {\mathcal F}_{A} \left( x,p \right) = 0.  
\end{equation} 
The classical Klimontovich microscopic phase density~\cite{Klimontovich-1995,Zubarev2,Liboff, Hakim} 
\begin{equation} \label{fA(x,p,t)} 
f_{A} \left(\mathbf{r},\mathbf{p},t\right) = \sum _{a} \ \delta^{3}\! \left( \mathbf{r}-\mathbf{r}_{aA} \left(t\right)\right) \, \delta^{3}\! \left( \mathbf{p}-\mathbf{p}_{aA} \left(t\right)\right) 
\end{equation} 
can be obtained from~\eqref{FA(x,p)} by integrating over the variable~$p^{0} $:
\begin{equation} \label{int-dp-1} 
	\begin{array}{c} 
		{\displaystyle \int\, dp^{0}\, p^{0}\, {\mathcal F}_{A} \left(x,p\right) } \\ 
		{\displaystyle=\int\, dp^{0}\, p^{0} \, \sum _{a} \, \delta^{3}\! \left( \mathbf{r}-\mathbf{r}_{aA} \left(t\right)\right) \, \delta^{3}\! \left( \mathbf{p}-\mathbf{p}_{aA} \left(t\right)\right) \, \delta \left(p^{0} -p^0_{aA} \right)\frac{m_{aA} }{p^0_{aA}} } \\ 
		{\displaystyle=m_{A}\, \sum _{a} \delta^{3}\! \left( \mathbf{r}-\mathbf{r}_{aA} \left(t\right)\right) \, \delta^{3}\! \left( \mathbf{p}-\mathbf{p}_{aA} \left(t\right)\right)  = m_{A} f_{A} \left(\mathbf{r,p},t\right).} 
	\end{array} 
\end{equation} 
It means that
\begin{equation} \label{FA2(x,p)} 
{  {\mathcal F}}_{A} \left(x,p\right)=\frac{1}{m_{A} p^{0} } \delta \left(p^{0} -\sqrt{\mathbf{p}^{2} +m_{A}^{2} c^{2} } \right)f_{A} \left(\mathbf{r,p},t\right). 
\end{equation} 
 In classical statistical mechanics, the interaction between particles is taken into account by introducing scalar functions --- potential energies. However, potential energy is an attribute of non-relativistic physics only. At the same time, at each point in space, we can determine the value of the scalar field $ \phi \left( \mathbf {r},t\right)$. 
A~smooth embedding of the Newtonian force $ - \nabla \phi \left( \mathbf {r},t\right)  $  in a pseudo-Euclidean space leads to the following 4-force~\cite{Kosyakov}:
\begin{equation} \label{Force} 
	F^{\mu }\left( x,p\right)  = \left(   g^{\mu \nu}- \frac{p^{\mu}p^{\nu}}{m^2c^2} \right) \, \frac{\partial \phi(x) }{\partial x^{\nu } },  
\end{equation} 
where $g^{\mu \nu}$ is the inverse of the metric tensor  in the Minkowski space.
The kinetic equation \eqref{dF-dx} will take the form
\begin{equation} \label{dF-dx_1} 
	\begin{array}{r}
		{\displaystyle \left( \frac{p^{\mu } }{m_{A} }\, \frac{\partial }{\partial x^{\mu } } +\left( g^{\mu \nu}- \frac{p^{\mu}\,p^{\nu}}{m_A^2\,c^2} \right)\, \frac{\partial \phi(x) }{\partial x^{\nu } }\, \frac{\partial  }{\partial p^{\mu } }  \right)  { {\mathcal F}}_{A} \left(x,p\right)}\\ \\
		{\displaystyle  = \frac {5p^{\mu}}{m_A^2\,c^2}\, \frac{\partial \phi(x) }{\partial x^{\mu } }\, { {\mathcal F}}_{A} \left( x,p \right). }
	\end{array}	
\end{equation} 
It should be emphasized that the equation~\eqref{dF-dx_1}  for particles of type $A$ contains \textit{not averaged} distribution functions~\eqref{FA(x,p)} and therefore this equation describes the dynamics of the system, not the dynamics of the probability density.

In the following, we will assume that there are no external forces. In this case, the 4-force~\eqref{Force} can be represented by the functional of the microscopic distribution function~\eqref{FA(x,p)}:\\
\begin{equation} \label{Force(F_A)} 
	F^{\mu }\left( x,p\right) = 	F^{\mu } \left( x, p,  \left\lbrace {\mathcal F}_{A}  \right\rbrace   \right).  
\end{equation}  

The equations~\eqref{dF-dx_1}--\eqref{Force(F_A)} form a closed system with respect to the distribution function~\eqref{FA(x,p)}. It is important to note that the value of the force~\eqref{Force(F_A)} at the point $(x,p)$ (that is, at some moment in time $t$) is determined by all distribution functions ${  {\mathcal F}}_{A} \left(x',p'\right)$, defined at the previous points $(x',p')$ of the Hilbert causality cone (that is, at the previous times $ t'<t $). This natural from a physical point of view causality principle from a mathematical point of view means that among all possible solutions of the equation of motion for the field through which the particles interact, we must exclude non-physical advanced solutions and keep the retarded solutions. The non-invariance of the kinetic equation~\eqref{dF-dx_1} with respect to time reversal
\begin{equation} \label{time_inver} 
	t\rightarrow-t, \ \mathbf{r} \rightarrow   \mathbf{r}, \ \mathbf{p} \rightarrow   -\mathbf{p}
\end{equation}
is related to taking into account \textit{only retarded potentials}.

To clarify the meaning of this statement, we write down the kinetic equation~\eqref{dF-dx_1} in terms of the Klimontovich distribution function~\eqref{fA(x,p,t)}. For this, we integrate the equation~\eqref{dF-dx_1} over $ p^{0} $ taking into account the relation~\eqref {FA2(x,p)}. As a result, we get:

\begin{equation} \label{df-dt_1} 
\begin{array}{c} 
{\displaystyle \left( \frac{\partial }{\partial t} 
		+\frac{c\mathbf{p}}{\sqrt{\mathbf{p}^{2} +m_{A}^{2} c^{2} } } \frac{\partial }{\partial \mathbf{r}} +\mathbf{F}\left(\mathbf{r,p},t\right)\frac{\partial }{\partial \mathbf{p}} \right) f_{A} \left(\mathbf{r,p},t\right) } \\
{=\displaystyle \frac{3}{m_{A}c^2}\left(\frac{\partial \phi(\mathbf{r},t)}{\partial t}+
	\frac{c\mathbf{p}}{\sqrt{\mathbf{p}^2+m_A^2c^2}}\frac{\partial \phi(\mathbf{r},t)}{\partial \mathbf{r}}\right)f_{A} \left(\mathbf{r,p},t\right)}.
\end{array} 
\end{equation} 

The formula for the force $ \mathbf{F}\left(\mathbf{r,p},t\right) $   follows from \eqref{Force} and the expression 
\begin{equation} \label{F_mu} 
F^{\mu } =\frac{p^0}{mc} \left( \frac{\mathbf{v} \mathbf{F}}{c} ,\mathbf{F} \right).
\end{equation} 
Thus, we have
\begin{equation} \label{F_} 
\begin{array}{c}
{\displaystyle \mathbf{F}\left(\mathbf{r,p},t\right) = -\frac{m_Ac}{\sqrt{\mathbf{p}^2+m_A^2c^2}} }\\ \\
{\displaystyle \times  \left[\frac {\partial }{\partial \mathbf{r}}+\frac{\mathbf{p}}{m_A^2c^2}\left(\mathbf{p}\frac{\partial }{\partial \mathbf{r}}+  \frac{\sqrt{\mathbf{p}^2+m_A^2c^2}}{c}\frac{\partial }{\partial t}\right)\right]\phi(\mathbf{r},t).  }
\end{array}	
\end{equation} 
In the instantaneously companion frame of reference we get classical expression for the Newtonian force:
\begin{equation} \label{force-New} 
\mathbf{F}(\mathbf{r},t)=-\frac{\partial \phi\left(\mathbf{r},t\right)}{\partial \mathbf{r}} .  
\end{equation} 
Note that by virtue of~\eqref{Force(F_A)} the scalar field $ \phi (x) $ is a linear functional of $ {\mathcal F}_{A} \left (x, p \right) $ and, therefore, of $ f_ {A} \left (\mathbf {r}, \mathbf {p}, t \right) $. According to the Riesz theorem on the representation of a linear continuous functional~\cite{Reed-1980}, the field $ \phi(x) $ can be written as follows:
\begin{equation}\label{Riesz}
	\begin{array}{c} 
	{\displaystyle \phi(\mathbf{r},t)= \iint dx'\, dp'\, K\left(x,p; x',p' \right) \, {\mathcal F}_{A} \left(x',p'\right) }.
	\end{array}	
\end{equation}

Let us show that the kernel $ {K} $ of the integral representation~\eqref{Riesz} is determined by a given interatomic potential. Indeed, the potential at point $\mathbf{r}$, taking into account the retardation of interaction $\tau _{a} $, can be written as
\begin{equation} \label{phi(r,t)1} 
\begin{array}{c} 
	{\displaystyle \phi\left(\mathbf{r},t\right)=\sum _{a}U\left(\mathbf{r-r}_{a} \left(t-\tau _{a} \right)\right) } \\ 
	{\displaystyle =\sum_{a}\int U\left(\mathbf{r-r}_{a} \left(t'\right)\right)  \delta \left(t-t'-\frac{\left|\mathbf{r-r}_{a} \right|}{c} \right)dt'} \\ 
	{\displaystyle =\iint U\left(\mathbf{r-r}'\right)  \delta \left(t-t'-\frac{\left|\mathbf{r-r}'\right|}{c} \right)\sum _{a}\delta \left(\mathbf{r}'-\mathbf{r}_{a} \left(t'\right)\right) d^{3} \mathbf{r}'\, dt',} 
\end{array} 
\end{equation} 
where $ U\left(\mathbf{r-r}'\right) $  is the potential energy determined for the particles \textit {at rest}, $c$~is the speed of transmission of the interaction (speed of light). Note that the resulting expression \eqref{phi(r,t)1} is an analog of the Li\'{e}nard–Wiechert potentials~\cite{Kosyakov}.

Integrating over time and introducing the dependence on the momentum, we get:
\begin{equation} \label{phi(r,t)2} 
\begin{array}{c} 
{\displaystyle \phi\left(\mathbf{r},t\right)=\int U\left(\mathbf{r-r}'\right) \sum_{a} \delta \left(\mathbf{r'-r}_{a} \left(t-\frac{\left|\mathbf{r-r}'\right|}{c} \right)\right) d^{3} \mathbf{r}'} \\ 
{\displaystyle =\iint U\left(\mathbf{r-r}'\right)  \sum _{a}\delta \left(\mathbf{r'-r}_{a} \left(t-\frac{\left|\mathbf{r-r}'\right|}{c} \right)\right) } \\ 
{\displaystyle \ \times \delta \left(\mathbf{p'-p}_{a} \left(t-\frac{\left|\mathbf{r-r}'\right|}{c} \right)\right)d^{3} \mathbf{r}'\, d^{3} \mathbf{p}'.} \end{array} 
\end{equation} 
As a result, we can write the potential \eqref{phi(r,t)2} in terms of the distribution function~\eqref{fA(x,p,t)}:
\begin{equation} \label{phi(r,t)3} 
\phi\left(\mathbf{r},t\right)=\iint U\left(\mathbf{r-r}'\right)  f_A\left(\mathbf{r',p}',t-\frac{\left|\mathbf{r-r}'\right|}{c} \right)d^{3} \mathbf{r}'d^{3} \mathbf{p}'. 
\end{equation} 
Comparing ~ \eqref{phi(r,t)3} with~\eqref{Riesz}, we conclude that the kernel of the linear operator~$ K\left(x,p; x',p' \right) $ has the form:
\begin{equation} \label{K} 
	K\left(x,p; x',p' \right)=\frac{m_A\, p'^{0}}{c}U\left(\mathbf{r-r}'\right)\,\delta \left(t-t'-\frac{\left|\mathbf{r-r}'\right|}{c}\right). 
\end{equation} 

For a qualitative analysis of the causes of irreversibility within the framework of the obtained kinetic equation~\eqref{df-dt_1}, we expand the distribution function in powers of the retardation time:
	\begin{equation} \label{decomp} 
	\begin{array}{c} 
		{\displaystyle f_A\left( \mathbf{r}',\mathbf{p}', t-{\frac{\left|\mathbf{r}-\mathbf{r}' \right|}{c}}\right) 
			} \\
		{\displaystyle =f_A\left( \mathbf{r}',\mathbf{p}',t \right)+ \sum_{s=1}^{\infty} \frac{(-1)^s}{s!}\left( {\frac{\left|\mathbf{r}-\mathbf{r}' \right|}{c}}\right)^{s} \frac{\partial^{s}\!  f_{A} \left( \mathbf{r}',\mathbf{p}', t\right) }{\partial t^s}}.
	\end{array} 
\end{equation} 

The time reversal operation~\eqref{time_inver} leads to a sign change only for members with odd values of $ s $. Since the equations~\eqref{df-dt_1},~\eqref{F_} and~\eqref{phi(r,t)3} contain terms that do not change sign under time inversion, the resulting kinetic equation is not invariant under the time inversion.\\

Thus, the kinetic equation~\eqref{df-dt_1} describes the irreversible evolution of a system of particles interacting via the scalar field $ \phi\left (\mathbf{r},t\right)$. Irreversibility is a consequence of the retardation of the interaction, which is expressed by the corresponding functional dependence of the potential on time.

\section{The law of energy change of a system of particles}
Let us multiply the equation of motion \eqref{eq-motion} by the delta function  $\delta ^{4}\! \left(x-x_{a} \left(\tau _{a} \right)\right)$, integrate over time $\tau _{a} $ and summate over $a$:
\begin{equation} \label{T-munu0} 
{\displaystyle \sum _{a}\int d\tau _{a}  \frac{dp_{a}^{\mu } }{d\tau _{a} }  \delta ^{4}\! \left(x-x_{a} \left(\tau _{a} \right)\right)=\sum _{a}\int d\tau _{a}  F_{a}^{\mu } \left(x_{a} ,p_{a} \right)\delta ^{4}\! \left(x-x_{a} \left(\tau _{a} \right)\right) }.  
\end{equation} 
Taking the integral on the left-hand side by parts, we get 
\begin{equation} \label{T-munu1} 
\partial _{\nu } T_{\mathrm{part}}^{\nu \mu} =c\sum _{a}\int d\tau _{a}  F_{a}^{\mu } \left(x_{a} ,p_{a} \right)\delta ^{4}\! \left(x-x_{a} \left(\tau _{a} \right)\right) .  
\end{equation} 
Here we have introduced the energy-momentum tensor for particles:
\begin{equation} \label{T-munu-2} 
T_{\mathrm{part}}^{\nu \mu} =\sum _{a}cm_{a}  \int d\tau _{a}\,  u_{a}^{\nu }\, u_{a}^{\mu }\, \delta ^{4}\! \left(x-x_{a} \left(\tau _{a} \right)\right). 
\end{equation} 
Integration over proper time brings the expression \eqref{T-munu1} to the form
\begin{equation} \label{d-nu-T} 
\partial _{\nu } T_{\mathrm{part}}^{\nu \mu} =\sum _{a}\gamma _{a}^{-1}  F_{a}^{\mu } \left(x_{a} ,p_{a} \right)\delta ^{3}\! \left(\mathbf{r-r}_{a} \left(t\right)\right). 
\end{equation} 
Note that 
\begin{equation} \label{T-00} 
 \begin{array}{c} 
 {\displaystyle T_{\mathrm{part}}^{00}  =\sum _{a}c\,m_{a}  \int d\tau _{a}\,  u_{a}^{0}\, u_{a}^{0}\, \delta ^{4}\! \left(x-x_{a} \left(\tau _{a} \right)\right)}\\
{\displaystyle =\sum _{a}m_{a}  u_{a}^{0}\, u_{a}^{0}\, \delta ^{3}\!\left(\mathbf{r-r}_{a} \right)\frac{d\tau _{a} }{dt} =\sum _{a}\delta ^{3}\! \left(\mathbf{r-r}_{a} \right)\gamma _{a}\, m_{a}\, c^{2}.} 
\end{array} 
\end{equation} 
is the energy density of the particles system.

Let us integrate \eqref{d-nu-T} over the volume occupied by the particle system
\begin{equation} \label{int-dnuT} 
\int d^{3} \mathbf{r}\,  \partial _{\nu } T_{\mathrm{part}}^{\nu \mu} =\int \sum _{a}\gamma _{a}^{-1}\, F_{a}^{\mu } \left(x_{a} ,p_{a} \right) \delta ^{3}\! \left(\mathbf{r-r}_{a} \left(t\right)\right)\, d^{3} \mathbf{r}. 
\end{equation} 
According to the Gauss theorem, the terms on the left-hand side of the equality containing divergence can be reduced to the flow of the energy-momentum tensor through the infinitely distant surface. Such integrals are equal to zero, since there are no charges and currents at infinity. As a result, taking into account $ \gamma cm = p^{0} $ we get
\begin{equation} \label{dE-dt} 
\frac{d}{dt} \sum _{a}m_{a} u_{a}^{\mu }  =c\int \sum _{a}\frac{m_{a} }{p^{0} } F_{a}^{\mu } \left(x_{a} ,p_{a} \right) \delta ^{3} \left(\mathbf{r-r}_{a} \left(t\right)\right) d^{3} \mathbf{r}.  
\end{equation} 
Assuming that the index $ \mu $ runs over the spatial values, we obtain the law of change in the total energy of the system of particles \eqref{T-00}: 
\begin{equation} \label{dE-dt2} 
\frac{d}{dt} \sum _{a}\frac{m_{a} c^{2} }{\sqrt{1-\left(\frac{\mathbf{v}_{a}}{c} \right)^{2} } }  =c^{2} \int \sum _{a}m_{a} \frac{F_{a}^{0} \left(x_{a} ,p_{a} \right)}{p^{0} }  \delta ^{3}\! \left(\mathbf{r-r}_{a} \left(t\right)\right) d^{3} \mathbf{r}.  
\end{equation} 
Since 
\begin{equation}\label{F-0}
 F^{0} =\gamma \frac{\left( \mathbf{v\cdot F}\right) }{c} =p^{0}\, \frac{\left( \mathbf{p\cdot F}\right)}{mc \sqrt{p^{2} +m^{2} c^{2} }}, 	
\end{equation}
the equation \eqref{dE-dt2} takes the form
\begin{equation} \label{dE-dt3} 
 \frac{d}{dt} \sum _{a}\frac{m_{a} c^{2} }{\sqrt{1-\left(\frac{\mathbf{v}_{a}}{c} \right)^{2} } }   =c\int \sum _{a}\frac{\mathbf{p}_{a} \mathbf{F}_{a} \left(\mathbf{r}_{a} ,\mathbf{p}_{a},t \right.)}{\sqrt{\mathbf{p}_{a}^{2} +m_{a}^{2} c^{2} } }  \delta ^{3} \left(\mathbf{r-r}_{a} \left(t\right)\right) d^{3} \mathbf{r}. 
\end{equation} 

Let us express the integral on the right-hand side in terms of the microscopic distribution function \eqref{fA(x,p,t)}:
\begin{equation} \label{dE-dt4} 
\begin{array}{l}
 { \displaystyle \iint \sum _{a}\frac{\mathbf{p}_{a} \mathbf{F}_{a} \left(\mathbf{r}_{a} ,\mathbf{p}_{a},t \right)}{\sqrt{p_{a}^{2} +m_{a}^{2} c^{2} } } \ \delta ^{3}\! \left(\mathbf{r-r}_{a} \left(t\right)\right)\delta ^{3}\! \left(\mathbf{p-p}_{a} \left(t\right)\right)  d^{3} \mathbf{r} d^{3} \mathbf{p}} \\ \\
  {\displaystyle =\iint \frac{\mathbf{pF}\left(\mathbf{r,p},t\right)}{\sqrt{p^{2} +m_{A}^{2} c^{2} } } \sum _{a} \delta ^{3}\! \left(\mathbf{r-r}_{a} \left(t\right)\right)\delta ^{3}\! \left(\mathbf{p-p}_{a} \left(t\right)\right)  d^{3} \mathbf{r}\, d^{3} \mathbf{p}}
   \\ \\
    {\displaystyle ={\iint}\frac{\mathbf{pF}\left(\mathbf{r,p},t\right)}{\sqrt{p^{2} +m_{A}^{2} c^{2} } }\ f_{A} \left(\mathbf{r,p},t\right)d^{3} \mathbf{r}\, d^{3} \mathbf{p} } 
 \end{array} 
\end{equation} 

Then, taking into account the expression \eqref{F_} for the force, we obtain the final expression for the rate of change of the total energy of the particles:
\begin{equation} \label{dE-dt5} 
\begin{array}{c} 
{\displaystyle  \frac{d}{dt} \sum _{a}\frac{m_{a} c^{2} }{\sqrt{1-\left(\frac{\mathbf{v}_{a}}{c} \right)^{2} } } } \\ 
{\displaystyle ={-\iint} \frac{f_{A} \left(\mathbf{r,p},t\right)}{m_A} \left(\mathbf{p}\frac {\partial }{\partial \mathbf{r}}+\frac{\mathbf{p}^2}{c\sqrt{p^{2} +m_{A}^{2} c^{2} }}\frac {\partial}{\partial t}\right) \phi(\mathbf{r},t)d^{3} \mathbf{r}d^{3} \mathbf{p} } \\
{\displaystyle ={-\iint} d^{3}\mathbf{r}d^{3} \mathbf{p}\frac{f_{A} \left(\mathbf{r,p},t\right)}{m_A} \left(\mathbf{p}\frac {\partial }{\partial \mathbf{r}}+\frac{\mathbf{p}^2}{c\sqrt{p^{2} +m_{A}^{2} c^{2} }}\frac {\partial}{\partial t}\right)} \\ 
{\displaystyle \times\iint U\left(\mathbf{r-r}'\right)  f_A\left(\mathbf{r',p}',t-\frac{\left|\mathbf{r-r}'\right|}{c} \right)d^{3} \mathbf{r}'d^{3} \mathbf{p}'.} \\
\end{array} 
\end{equation} 
From this, as well as from the equation \eqref{df-dt_1} taking into account the expansion \eqref{decomp}, it follows that the total energy of the system of particles changes irreversibly.

\section{Evolution of a system of particles in the first approximation with respect to the retardation time}

In the first approximation in the time retardation, we have
\begin{equation} \label{f-ret-1} 
f\left(\mathbf{r',p}',t-\frac{\left|\mathbf{r-r}'\right|}{c} \right)=f\left(\mathbf{r',p}',t\right)-\frac{\left|\mathbf{r-r}'\right|}{c} \frac{\partial }{\partial t} f\left(\mathbf{r',p}',t\right).  
\end{equation}  
Taking into account \eqref {f-ret-1}, leaving in the expansion the terms of order not higher than $ c^{- 1} $, the kinetic equation \eqref{df-dt_1} can be represented as:
\begin{equation} \label{f-ret-2} 
	\begin{array}{c} 
		{\displaystyle \frac{\partial f_{A} \left(\mathbf{r,p},t\right)}{\partial t} +\frac{\mathbf{p}}{m_{A} } \frac{\partial f_{A} \left(\mathbf{r,p},t\right)}{\partial \mathbf{r}} } \\ 
		{\displaystyle =\frac{\partial f_{A} \left(\mathbf{r,p},t\right)}{\partial \mathbf{p}} \Bigg[ {\iint}\frac{\partial U\left(\mathbf{r-r}'\right)}{\partial \mathbf{r}} f_{A} \left(\mathbf{r',p'},t\right)d^{3} \mathbf{r}'d^{3} \mathbf{p}'} \\  
		{\displaystyle -\iint \left(U\left(\mathbf{r-r}'\right)\frac{\mathbf{r-r}'}{\left|\mathbf{r-r}'\right|} +\frac{\partial U\left(\mathbf{r-r}'\right)}{\partial \mathbf{r}} \left|\mathbf{r-r}'\right|\right)  } \\ 
		{\displaystyle \times \frac{1}{c}\, \frac{\partial f_{A} \left(\mathbf{r',p}',t\right)}{\partial t}   d^{3}\mathbf{r}'d^{3} \mathbf{p'}\Bigg].} 
	\end{array} 
\end{equation} 
Since in the case of an arbitrary potential $ U \left(\mathbf{r} \right) $ the quantity 
	\begin{equation}\label{U(r)-gen}
	W\left( \mathbf{r} \right) =	\left(U\left(\mathbf{r}\right)\frac{\mathbf{r}}{\left|\mathbf{r}\right|} +\frac{\partial U\left(\mathbf{r}\right)}{\partial \mathbf{r}} \left|\mathbf{r}\right|\right) \not\equiv 0,
	\end{equation}
then already in the first order in retardation the equation~\eqref{f-ret-2} is irreversible.

There is the only exception:  the Coulomb potential $ U_{\mathrm {Coul}} (\mathbf{r}) \sim \left| \mathbf {r} \right|^{- 1} $, for which $ W_{\mathrm {Coul}} (\mathbf {r}) \equiv 0 $ and irreversibility manifests itself only when higher order terms are taken into account. The same holds for the Lorentz force:: irreversibility is manifested starting from the third-order terms $ c^{-3}$. This fact has far-reaching consequences.

Let us introduce the notion of the moment of the potential:
\begin{equation} \label{f-ret-3} 
U_{\mathrm{mom} } \left(\mathbf{r-r}'\right)=U\left(\mathbf{r-r}'\right) {\left|\mathbf{r-r}'\right|}.  
\end{equation} 
Then we have
\begin{equation} \label{dU-dr1} 
\left(U\left(\mathbf{r-r}'\right)\frac{\mathbf{r-r}'}{\left|\mathbf{r-r}'\right|} +\frac{\partial U\left(\mathbf{r-r}'\right)}{\partial \mathbf{r}} \left|\mathbf{r-r}'\right|\right) =\frac{\partial }{\partial \mathbf{r}} U_{\mathrm{mom}} \left(\mathbf{r-r}'\right),  
\end{equation} 
and the rate of change in the total energy of the system of particles in the first approximation in the retardation time can be represented in the form
\begin{equation} \label{d--dt-E} 
\begin{array}{c} {\displaystyle \frac{d}{dt} \sum _{a}\frac{m_{a} c^{2} }{\sqrt{1-\left({v_{a} \mathord{\left/ {\vphantom {v_{a}  c}} \right. \kern-\nulldelimiterspace} c} \right)^{2} } }  } \\ \\ {\displaystyle =- \iint d^{3}\mathbf{r}\, d^{3}\mathbf{p}\ f_{A} \left(\mathbf{r,p},t\right)\frac{\mathbf{p}}{m_{A} } \iint \frac{\partial U\left(\mathbf{r-r}'\right)}{\partial \mathbf{r}}\ f_{A} \left(\mathbf{r',p}',t\right) d^{3}\mathbf{r'}\, d^{3}\mathbf{p'}  } \\ \\ 
	{\displaystyle  +{\iint}d^{3}\mathbf{r}\, d^{3}\mathbf{p}\ f_{A} \left(\mathbf{r,p},t\right)\ \frac{\mathbf{p}}{m_{A} }  } \\ \\ 
	{\displaystyle \times {\iint}\frac{\partial U_\mathrm{mom} \left(\mathbf{r-r}'\right)}{\partial \mathbf{r}}\ \frac{1}{c} \frac{\partial f_{A} \left(\mathbf{r',p'},t\right)}{\partial t}\, d^{3}\mathbf{r'}\, d^{3}\mathbf{p'}.} 
\end{array} 
\end{equation} 
Let us take into account that in the first approximation in $v_{a}/{c}  $ the total energy of particles can be written as
\begin{equation}\label{E-kin}
	\sum _{a}\frac{m_{a} c^{2} }{\sqrt{1-\left({v_{a} \mathord{\left/ {\vphantom {v_{a}  c}} \right. \kern-\nulldelimiterspace} c} \right)^{2} } } \approx  \sum _{a}\left(m_{a}c^2 + \frac{m_{a} v_{a}^{2} }{2}\right).
\end{equation}
The first term on the right-hand side of the equation \eqref{d--dt-E} is the derivative of the potential interaction energy of \textit{resting} point particles (or the potential energy of a system of interacting particles in the non-relativistic approximation). Indeed
\begin{equation} \label{d--dt--E} 
\begin{array}{c} 
{\displaystyle \int d^{3} \mathbf{r} \int d^{3} \mathbf{p}\, f_{A} \left(\mathbf{r,p},t\right)\frac{\mathbf{p}}{m_{A} } \frac{\partial }{\partial \mathbf{r}} \int d^{3} \mathbf{r}' \int d^{3} \mathbf{p}'\, U\left(\mathbf{r-r}'\right)f_{A} \left(\mathbf{r',p'},t\right)} \\ \\ 
{\displaystyle =\frac{1}{2} \int d^{3} \mathbf{r} \int d^{3} \mathbf{p} \, f_{A} \left(\mathbf{r,p},t\right) } \\ \\ 
{\displaystyle \times  \frac{\partial }{\partial \mathbf{r}} \int d^{3} \mathbf{r'} \int d^{3} \mathbf{p'}\, \frac{\left(\mathbf{p-p}'\right)}{m_{A}} \,  U\left(\mathbf{r-r'}\right)\, f_{A} \left(\mathbf{r',p'},t\right)} \\ \\
{\displaystyle =\frac{d}{dt}\left(  \frac{1}{2} \sum _{a}\sum _{b}U_{ab}   \left(\mathbf{r}_{a}\left( t\right) -\mathbf{r}_{b}\left( t\right) \right)\right) \equiv \frac{d}{dt} U\left( t\right).} 
\end{array} 
\end{equation} 
Thus, the equation~\eqref{d--dt--E} can be written in the form
\begin{equation}\label{1-st-corr}
\begin{array}{c}
	{\displaystyle  \frac{d}{dt} \left(\sum _{a}\frac{m_{a} v_{a}^{2}\left( t\right) }{2}  +\frac{1}{2} \sum _{a}\sum _{b}U_{ab}   \left(\mathbf{r}_{a}\left( t\right) -\mathbf{r}_{b}\left( t\right) \right)\right) }\\
	{\displaystyle  =  {\iint}d^{3}\mathbf{r}\, d^{3}\mathbf{p}\   f_{A} \left(\mathbf{r,p},t\right)\ \frac{\mathbf{p}}{m_{A} }\ }\\
	{\displaystyle \times {\iint}\frac{\partial U_\mathrm{mom} \left(\mathbf{r-r}'\right)}{\partial \mathbf{r}}\ \frac{1}{c} \frac{\partial f_{A} \left(\mathbf{r',p'},t\right)}{\partial t}\, d^{3}\mathbf{r'}\, d^{3}\mathbf{p'}.}
\end{array}	
\end{equation}
Hence it follows that the first relativistic correction in the equation~\eqref{f-ret-2} describes the change in the non-relativistic energy of a system of interacting particles.

Consider now the second term on the right-hand side of the formula \eqref{d--dt-E}. For this, we take into account the following identities
\begin{equation} \label{j(r,t)-full} 
\int d^{3}\mathbf{p} \, \frac{\mathbf{p}}{m_{A} }\, f_{A} \left(\mathbf{r,p},t\right)=\sum _{a}\frac{\mathbf{p}_{a} }{m_{A} }  \delta \left(\mathbf{r-r}_{a} \left(t\right)\right)=\mathbf{j}\left(\mathbf{r},t\right), 
\end{equation} 
\begin{equation} \label{divj(r,t)} 
\begin{array}{c}
{\displaystyle \int \frac{\partial f_{A} \left(\mathbf{r',p'},t\right)}{\partial t}\,  d^{3} \mathbf{p}'=\frac{\partial }{\partial t} \int f_{A} \left(\mathbf{r',p}',t\right) d^{3} \mathbf{p'}}\\
{\displaystyle =\frac{\partial }{\partial t} n\left(\mathbf{r}',t\right)=-{ \mathrm{ div}}\, \mathbf{j}\left(\mathbf{r}',t\right), }
\end{array}
\end{equation} 
where $ \mathbf{j}\left(\mathbf{r}, t \right) $ is the vector of the particle flux density. 
As a result, the second term in \eqref{d--dt-E} takes the form
\begin{equation} \label{int-rp-f} 
\begin{array}{c} 
	{\displaystyle  {\iint}d^{3}\mathbf{r}\, d^{3}\mathbf{p}\ f_{A} \left(\mathbf{r,p},t\right) \frac{\mathbf{p}}{m_{A} }  } \\ \\ 
	{\displaystyle \times {\iint}\frac{\partial U_\mathrm{mom} \left(\mathbf{r-r}'\right)}{\partial \mathbf{r}}\ \frac{1}{c}\frac{\partial f_{A} \left(\mathbf{r',p'},t\right)}{\partial t}\, d^{3}\mathbf{r'}\, d^{3}\mathbf{p'}} \\ \\
 {\displaystyle =-\frac{1}{c}\int d^{3} \mathbf{r}\ \mathbf{j}\left(\mathbf{r},t\right)\int \frac{\partial U_\mathrm{mom} \left(\mathbf{r-r}'\right)}{\partial \mathbf{r}}\  {  \mathrm{div}}\, \mathbf{j}\left(\mathbf{r'},t\right)d^{3} \mathbf{r}'} \\ \\
  {\displaystyle =\frac{1}{c}\int d^{3} \mathbf{r}\ j_{\alpha } \left(\mathbf{r},t\right) \int \frac{\partial ^{2} U_\mathrm{mom} \left(\mathbf{r-r}'\right)}{\partial x_{\alpha } \partial x'_{\beta } }\, j_{\beta } \left(\mathbf{r}',t\right)d^{3} \mathbf{r}' .} 
\end{array} 
\end{equation} 

Using the representation of the moment of potential in the form of the Fourier integral
\begin{equation} \label{U-mom} 
U_\mathrm{mom} \left(\mathbf{r-r}'\right)=\frac{1}{\left(2\pi \right)^{3} } \int d^{3} \mathbf{k}\, \tilde{U}_\mathrm{mom} \left(\mathbf{k}\right)\exp \left(-i\mathbf{k}\left(\mathbf{r-r}'\right)\right),  
\end{equation} 
we find
\begin{equation} \label{dU-mom} 
\frac{\partial ^{2} U_\mathrm{mom}\left(\mathbf{r-r}'\right)}{\partial x_{\alpha } \partial x'_{\beta } } =\frac{1}{\left(2\pi \right)^{3} } \int d^{3} \mathbf{k}\, \tilde{U}_\mathrm{mom} \left(\mathbf{k}\right)k_{\alpha } k_{\beta } \exp \left(-i\mathbf{k}\left(\mathbf{r-r}'\right)\right).  
\end{equation} 
Consequently, the second term on the right-hand side of the formula~\eqref{d--dt-E} has the following form
\begin{equation} \label{int-j-U-mom} 
\begin{array}{c} 
{\displaystyle \frac{1}{\left(2\pi \right)^{3} } \int d^{3} \mathbf{k}\, \tilde{U}_\mathrm{mom} \left(\mathbf{k}\right)k_{\alpha } k_{\beta } } \\ 
{\displaystyle \times \int d^{3} \mathbf{r}\,  j_{\alpha } \left(\mathbf{r},t\right)\exp \left(-i\mathbf{kr}\right) \int d^{3} \mathbf{r'}\, j_{\beta } \left(\mathbf{r'},t\right) \exp \left(i\mathbf{kr'}\right) } \\ 
{\displaystyle =\frac{1}{\left(2\pi \right)^{3} } \int d^{3} \mathbf{k}\left|\mathbf{k}\tilde{\mathbf{j}}\left(\mathbf{k},t\right)\right|^{2}  \tilde{U}_\mathrm{mom} \left(\mathbf{k}\right).} 

\end{array} 
\end{equation} 
As a result, the rate of change in the particle energy in the first approximation in the retardation of interactions has the following form:
\begin{equation} \label{int-j-U-mom-2} 
	\begin{array}{c}
{\displaystyle \frac{d}{dt} \left(\sum _{a}\frac{m_{a} v_{a}^{2}\left( t\right) }{2}  +\frac{1}{2} \sum _{a}\sum _{b}U_{ab}   \left(\mathbf{r}_{a}\left( t\right) -\mathbf{r}_{b}\left( t\right) \right)\right) }\\
	{\displaystyle =\frac{1}{\left(2\pi \right)^{3}c } \int d^{3} \mathbf{k}\left|\mathbf{k}\tilde{\mathbf{j}}\left(\mathbf{k},t\right)\right|^{2}  \tilde{U}_\mathrm{mom} \left(\mathbf{k}\right).}
\end{array}
\end{equation} 
Thus, the evolution of the non-relativistic energy of a system of interacting particles in the first approximation in the retartation of interactions is determined by the Fourier transform of the moment of the interparticle potential.

Let us express  $ \tilde{U}_\mathrm{mom} \left(\mathbf{k}\right) $ in terms of the Fourier transforms of the interatomic potential 
${U}\left(\mathbf{r}\right)$ and the function $V \left( \mathbf{r}\right)  = \left| \mathbf{r} \right|.$
The Fourier transform of the function $ V \left (r \right) $, due to its singularity in the vicinity of the point at infinity, exists only in the class of generalized functions (distributions) and has the form~\cite{Gelfand}:
\begin{equation}\label{W(k)}
	\tilde{V}(\mathbf{k}) = -\frac{8\pi}{\left| \mathbf{k}\right|^{4} }.
\end{equation}
By the definition of ~\eqref{f-ret-3}, the Fourier transform of the moment of a potential is the convolution of the functions $ \tilde {U} \left (\mathbf {k} \right) $ and $ \tilde {V} (\mathbf {k}) $:
\begin{equation}\label{pot-to-mom}
\tilde{U}_\mathrm{mom} \left(\mathbf{k}\right)=-\frac{1}{\pi^{2}} \int  d^{3} \mathbf{q}\ \frac{\tilde{U} \left( \mathbf{q}\right) } {\left|\mathbf{k - q}\right|^{4} }.	
\end{equation}  

Substitute this expression into the formula~\eqref{int-j-U-mom-2} and find
\begin{equation}\label{int-jUmom}
	\begin{array}{c}
	{\displaystyle \frac{d}{dt} \left(\sum _{a}\frac{m_{a} v_{a}^{2}\left( t\right) }{2}  +\frac{1}{2} \sum _{a}\sum _{b}U_{ab}   \left(\mathbf{r}_{a}\left( t\right) -\mathbf{r}_{b}\left( t\right) \right)\right) }\\
	{\displaystyle =-\frac{4}{\left(2\pi \right)^{5}c } \int d^{3} \mathbf{k}\left|\mathbf{k}\tilde{\mathbf{j}}\left(\mathbf{k},t\right)\right|^{2}  \int  d^{3} \mathbf{q}\ \frac{\tilde{U} \left( \mathbf{q}\right) } {\left|\mathbf{k - q}\right|^{4} }.}
\end{array}	
\end{equation}
The resulting formula establishes a connection between the evolution of the total energy of a system with retarted interactions and the characteristics of interatomic interactions.

\section{The stability criterium of interatomic potentials and the zeroth law of thermodynamics}
Inter-atomic interactions in condensed matter physics are described, as a rule, using various model potentials. An extensive literature is devoted to the problem of finding inter-atomic potentials~\cite{Vaks, Finnis, Kaplan, Israelachvili}.

We will not limit our consideration to the choice of a specific form of model inter-atomic potentials. Let us take into account the only significant limitation on the explicit form of inter-atomic interactions. This limitation is due to the requirement for the existence of a thermodynamic limit, according to which the logarithm of the partition function of any system must be an extensive function. Inter-atomic potentials satisfying this requirement are called stable or non-catastrophic potentials. The criterion for the stability of inter-atomic potentials was established in the works of Dobrushin, Fisher and Ruelle~\cite{Dob-1, Fish-1, Fisher-Rue, Ruelle}. In terms of the Fourier transform of potentials for the case of two-body inter-atomic potentials, this criterion has the form~\cite {Baus}:

\begin{equation}\label{non-neg}
	\tilde{U}\left( \mathbf{q}\right) \geq 0.
\end{equation}
Thus, the function $ \tilde {U}_\mathrm{mom} \left (\mathbf {k} \right) $~\eqref {pot-to-mom} is non-positive for all $\mathbf{k} $. Therefore, we have:
	\begin{equation}\label{int-jUmom2}
		\begin{array}{c}
			{\displaystyle \frac{d}{dt} \left(  \sum _{a}\frac{m_{a} v_{a}^{2}\left( t\right)  }{2}  +\frac{1}{2} \sum _{a}\sum _{b}U   \left(  \mathbf{r}_{a}\left( t\right)  -\mathbf{r}_{b}\left( t\right)  \right)  \right) }\\
			{\displaystyle =-\frac{4}{\left(2\pi \right)^{5}c } \int d^{3} \mathbf{k}\left|\mathbf{k}\tilde{\mathbf{j}}\left(\mathbf{k},t\right)\right|^{2}  \int  d^{3} \mathbf{q}\ \frac{\tilde{U} \left( \mathbf{q}\right) } {\left|\mathbf{k - q}\right|^{4} } \leq 0.}
		\end{array}	
	\end{equation}
	
Energy constancy is possible if and only if
\begin{equation}\label{j=0}
	\tilde{\mathbf{j}}\left(\mathbf{k},t\right) \equiv 0 	
\end{equation}
for all $ \mathbf{k} $, i.e. all particles are at rest.

\section{Discussion and future directions}

The main results of this work are as follows.	
	\begin{enumerate}
		\item A closed classical relativistic dynamical theory of a system of identical particles interacting with each other through a scalar field is constructed.
		
		\begin{itemize}
			\item 
			The dynamics of a system of particles is described by the kinetic equation ~\eqref{dF-dx_1} in terms of the relativistically invariant microscopic distribution function ~\eqref{FA(x,p)}.
			\item 
			The dynamic scalar field through which the particles interact is expressed through the interatomic potentials of the particles at rest and the microscopic distribution functions ~\eqref{Riesz},~\eqref{phi(r,t)3}.
		\end{itemize}
		
		\item 
		It has been established that for the class of inter-atomic potentials stable according to Dobrushin-Fisher-Ruelle, the system of interacting particles, regardless of its initial state, in the absence of external fields, irreversibly passes into a state with minimum energy.
		\item 
		The principles of relativity and the retardation of interactions associated with the principle of causality are sufficient for the existence of a state of thermodynamic equilibrium in both many-body and few-body isolated classical systems.
		
		\end{enumerate}
	
	Thus, the relativity theory and the principle of causality lead to the following conclusions.
	
	\begin{enumerate}
		\item 
		The impossibility of instant interactions between particles is the reason that generates the thermodynamic behavior of systems.
		\item  		
		A consistent microscopic explanation and substantiation of the zeroth law of thermodynamics within the framework of classical non-relativistic mechanics is impossible.
		\item 
		In the absence of external influences, the energy of a system of particles, the interaction potential between which at rest satisfies the Dobrushin-Fisher-Ruelle stability criterion, monotonically decreases over time and tends to a minimum. All particles are at rest in this state.
		\item 
		The energy excess of the initial state over the final state of the system is carried away by the field through which the particles interact.
		
	\end{enumerate}
	
	What is the state of thermodynamic equilibrium of real systems?

Within the framework of the concept developed in this work, a more thorough analysis of the notion of an isolated system of particles is required. This is due to the fact that the concept of the potential energy of a system of interacting particles, which depends on the instantaneous simultaneous coordinates of these particles, does not exist within the framework of the relativistic theory. Therefore, the field through which the particles interact begins to play a decisive role. In fact, the field is an additional and very specific component of the system.
	
	Within the framework of classical thermodynamics, an isolated system is a system of interacting particles that does not exchange either matter or energy with the surrounding world. In the relativistic theory, the issue of exchange with the outside world is considered separately for particles and for the field.\\
	
	In this regard, we will consider two options for placing a particle system in a box.
	\begin{itemize}
			\item 
		The walls are impenetrable to particles, but are transparent for  the field. In this case, the escaping field carries away the excess of energy, and the system of particles inside the box comes to the ground state, in which all particles are at rest in accordance with the equation~\eqref{j=0}.\\
		
		\item 
		The walls are impenetrable to both particles and the field. Then all the particles of the system are under the influence of the ``external'' field reflected by the walls of the box. In this case, one can expect that the system of particles passes into a stationary state of dynamic equilibrium with the field, by analogy with the simple model~\cite{Zakharov21-1}.
		
	\end{itemize}

Since the interactions between atoms are of an electromagnetic nature and all real systems are always immersed in an alternating intrinsic and external electromagnetic field, the exchange of energy between macroscopic bodies to a large extent occurs not directly, but through a universal mediator -- the electromagnetic field. The retardation of interactions transmitted by an electromagnetic field is a universal mechanism that implements the laws of thermodynamics.

\section{Acknowledgements}

The authors are grateful to Prof. Ya.I. Granovsky and Prof. V.V. Uchaikin for lively inspiring discussions and constructive criticism.

\end{document}